\documentclass[twocolumn,superscriptaddress,nofootinbib,floatfix,aps,prb,citeautoscript,longbibliography]{revtex4-2}
\usepackage[utf8]{inputenc}
\usepackage[T1]{fontenc}
\usepackage{lineno}

\usepackage[colorlinks=true,linkcolor=blue,citecolor=blue,urlcolor=blue]{hyperref}
\usepackage{graphicx}
\usepackage{xcolor}
\usepackage{xstring}
\usepackage{array}
\usepackage{dcolumn}
\usepackage{booktabs}
\usepackage{bm}
\usepackage{multirow}
\usepackage[version=4]{mhchem}
\usepackage{cleveref}
\usepackage{orcidlink}
\usepackage{float}

\date{\today}

\usepackage{soul}
\sethlcolor{yellow}

\begin{document}

\newcommand{\bochum}{Research Center Future Energy Materials and Systems of the University Alliance Ruhr and Interdisciplinary Centre for Advanced Materials Simulation, Ruhr University Bochum, Universit\"atsstraße 150, D-44801 Bochum, Germany}
\newcommand{\RUBPhysics}{Faculty of Physics and Astronomy, Ruhr University Bochum, Universit\"atsstraße 150, D-44801 Bochum, Germany}
\newcommand{\coimbra}{CFisUC, Department of Physics, University of Coimbra, Rua Larga, 3004-516 Coimbra, Portugal}
\newcommand{\rome}{Dipartimento di Fisica, Sapienza University of Rome, Piazzale Aldo Moro 5, 00185 Roma, Italy}

\newcommand{\new}[1]{\textcolor{black}{#1}}
\newcommand{\miguel}[1]{\textcolor{blue}{#1}}
\newcommand{\haichen}[1]{\textcolor{green}{#1}}
\newcommand{\thalis}[1]{\textcolor{red}{#1}}

\author{Thalis H. B. da Silva}
\affiliation{\bochum}

\author{Hai-Chen Wang}
\affiliation{\bochum}

\author{Tiago F. T. Cerqueira}
\affiliation{\coimbra}

\author{Simone Di Cataldo}
\affiliation{\rome}

\author{Silvana Botti}
\email{silvana.botti@rub.de}
\affiliation{\bochum}
\affiliation{\RUBPhysics}

\author{Miguel A. L. Marques}
\email{miguel.marques@rub.de}
\affiliation{\bochum} 

\title{High-throughput study of electrical conductivity in ordered metals}

\begin{abstract}
\noindent\textbf{Abstract}\\[0.5em]
We present a computational framework that integrates machine learning with high-throughput \textit{ab initio} calculations to screen over 2.8 million compounds for metallic transport. We identify several intermetallic candidates with predicted high conductivities comparable to that of aluminum ($36.59 \times 10^6$~S/m). We perform full electron--phonon coupling calculations for the top-performing materials, yielding results in good agreement with available experimental data. Our analysis reveals that while the noble metals (Ag, Au, Cu) possess a conductivity that remains difficult to surpass due to their unique electronic structure and low scattering, compounds like $\text{LiBePt}_2$ can achieve comparable performance by utilizing valence electrons from light elements to shift high-scattering $d$-states beneath the Fermi level. This study not only identifies novel high-performance conductors but also demonstrates the predictive power of combining statistical learning with detailed ab initio calculations.
\end{abstract}

\maketitle

\section*{Introduction}

Electrical conductivity is one of the most fundamental and technologically important properties of metallic materials, governing their applications across diverse fields from microelectronics to power transmission, energy storage, and quantum technologies. The ability to predict and optimize electrical transport properties from first principles has long been a central goal in computational materials science, as it enables the rational design of materials with tailored electronic properties without the need for extensive experimental synthesis and characterization.

\new{Traditional approaches to calculating electrical conductivity in metals rely on density functional theory (DFT) combined with Boltzmann transport theory, which provides a quantum mechanical framework for understanding electron transport phenomena~\cite{Scheidemantel2003, Bernardi2016, Ponc2020}. This semiclassical description assumes well-defined quasiparticles, an assumption that breaks down beyond the weak electron-phonon coupling limit. More rigorous formulations based on the Kubo formula~\cite{Kubo1957, Allen2015} have recently been implemented from first principles, ranging from the bubble approximation, which retains the full electronic spectral functions~\cite{Zhou2019, Lihm2025}, to ladder-type resummations that additionally include vertex corrections~\cite{Lihm2026}; such beyond-quasiparticle approaches have been applied to correlated metals such as SrVO$_3$~\cite{Abramovitch2024}, albeit at a substantially higher computational cost. Indeed, already at the semiclassical level these methods are computationally intensive, typically requiring detailed calculations of electronic band structures, scattering rates, and transport integrals that limit their application to small sets of materials. While significant advances have been made in computational efficiency and the development of specialized codes for transport calculations~\cite{epw0, epw1, epw2, epw3}, the exploration of vast chemical spaces remains prohibitively expensive using conventional DFT approaches alone.}

The emergence of high-throughput computational materials discovery has revolutionized the field by enabling systematic screening of millions of compounds for desired properties. Databases, such as the Materials Project~\cite{MP}, AFLOW~\cite{Aflowlib}, OQMD~\cite{OQMD}, and Alexandria~\cite{alex1, alex2,newalexpaper} have catalogued electronic, structural, and thermodynamic properties of hundreds of thousands of materials. However, transport properties like electrical conductivity have remained largely underexplored in these high-throughput frameworks due to their computational complexity and the additional approximations required beyond standard DFT calculations.

Recent developments in machine learning (ML) have opened opportunities to accelerate materials discovery by learning complex structure-property relationships from existing computational or experimental data~\cite{cgat, alignn, crabnet, modnet}. ML models can potentially reduce the computational cost of property predictions by several orders of magnitude while maintaining reasonable accuracy, making it feasible to screen millions of materials that would be computationally intractable using traditional methods. 
For electrical conductivity, this approach is particularly promising as it can capture the intricate relationships between crystal structure, electronic structure, and transport properties that govern metallic behavior.

\new{In this work, we employ DFT together with Boltzmann transport equation (BTE) calculations to generate a reliable dataset of transport-related quantities, where the linearized BTE is solved iteratively (iBTE).} We present a comprehensive high-throughput investigation of electrical conductivity in metallic systems, that allows us to screen approximately 2.8 million inorganic ordered compounds. Our approach follows a multi-stage workflow in which the Eliashberg parameter $\lambda$, previously computed in the context of conventional superconductivity studies, is incorporated as an additional descriptor.

We validate the accuracy and robustness of this framework through extensive comparisons with both computational results and available experimental data for electrical conductivities.

Furthermore, our results demonstrate the predictive power of the BTE formalism for metallic conductivity. The calculated electrical conductivities reproduce experimental values with relative errors below $30\%$ for the majority of systems considered.
\section*{Results}

\label{sec:res}
\subsection*{Machine Learning Screening}

\begin{figure}[tbh]
\centering
\includegraphics[width=0.99\columnwidth]{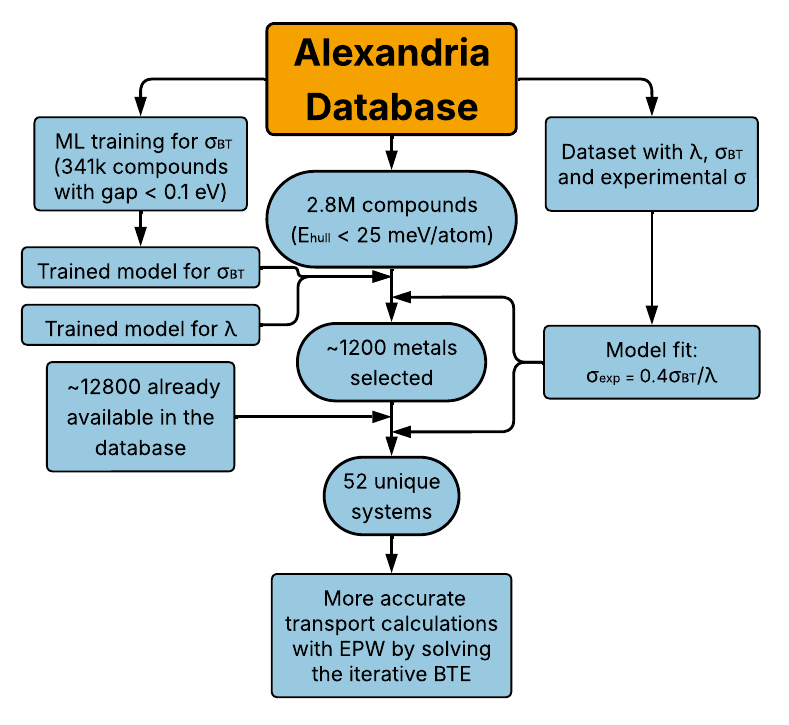}
\caption{\new{Schematic overview of the computational workflow. Starting from the ${\sim}2.8$~M compounds in the Alexandria database with $E_{\mathrm{hull}} < 25$~meV/atom, two ML models (a pre-existing one for the electron-phonon coupling $\lambda$ and one trained here for the BoltzTraP conductivity $\sigma_{\mathrm{BT}}$) predict $\lambda$ and $\sigma_{\mathrm{BT}}$ for the full set. An empirical fit, $\sigma_{\mathrm{SR}} = 0.4\,\sigma_{\mathrm{BT}}/\lambda$, calibrated against experimental conductivities, then estimates $\sigma_{\mathrm{exp}}$, and compounds exceeding $15\times10^{6}$~S/m are retained (${\sim}1200$~metals). These undergo explicit BoltzTraP and electron-phonon calculations and are merged with ${\sim}12{,}800$ compounds already available in the database (${\sim}14{,}000$ total). Reapplying the same fit narrows this set to 52 unique systems, for which accurate transport properties are finally computed with EPW by solving the iterative Boltzmann transport equation.}}
\label{fig:flowchart}
\end{figure}

Obtaining quantitatively reliable transport properties from first-principles theory requires computationally demanding calculations.
To circumvent this bottleneck, we employ a machine-learning-accelerated workflow, depicted schematically in Fig.~\ref{fig:flowchart}, in which two physically motivated descriptors serve as computationally efficient proxies for the electrical conductivity. The first is the electrical conductivity computed using BoltzTraP2 within the constant relaxation-time approximation~\cite{boltztrap2}. The second is the electron--phonon coupling constant $\lambda$ from Eliashberg theory, which encodes the strength of the dominant scattering mechanism limiting conductivity in crystalline metals. Machine learning surrogate models are trained to predict both descriptors efficiently, and their predictions are combined through an empirical formula derived via symbolic regression to screen candidate high-conductivity metals. All training data and screened compounds are from the Alexandria database~\cite{alex1, alex2}.

The electrical conductivity computed by BoltzTraP2~\cite{boltztrap2} depends solely on the electronic band structure, making it relatively straightforward and computationally inexpensive to evaluate in a high-throughput framework. However, this approach uses the constant relaxation-time approximation and therefore does not explicitly account for electron--phonon scattering processes, which are the dominant mechanism limiting charge transport in metals. As a consequence, $\sigma_{\mathrm{BT}}$ cannot be directly compared with experimental conductivities $\sigma$, since material-specific variations in scattering strength are entirely neglected. To remedy this deficiency, we propose to incorporate the electron--phonon coupling constant $\lambda$ from Eliashberg theory as a proxy for the scattering strength. This choice is physically motivated by the close formal analogy between the transport and Eliashberg definitions of $\lambda$: both quantities are expressed as Fermi-surface averages of the electron--phonon matrix elements, differing only in their respective weighting kernels as shown in the equations below~\cite{epw1, epw2}.

\begin{widetext}
\begin{equation}
    \lambda^\text{tr}_{\bm{q}\nu} =  \frac{1}{N(E_\text{F})\omega_{\bm{q}\nu}}
    \sum_{n,m}\int_\text{BZ} \frac{d^3k}{V_\text{BZ}} 
    \left|g_{mn,\nu}(\bm{k}, \bm{q})\right|^2\delta(\varepsilon_{n\bm{k}} - E_\text{F})\delta(\varepsilon_{m\bm{k+q}} - E_\text{F})
    \left(1 - \frac{\bm{v}_{n\bm{k}} \cdot \bm{v}_{m\bm{k+q}}}{|v_{n\bm{k}}| |v_{m\bm{k+q}}|}\right)
    \,,
\end{equation}

\begin{equation}
    \lambda_{\bm{q}\nu} =  \frac{1}{N(E_\text{F})\omega_{\bm{q}\nu}}
    \sum_{n,m}\int_\text{BZ} \frac{d^3k}{V_\text{BZ}} 
    \left|g_{mn,\nu}(\bm{k}, \bm{q})\right|^2\delta(\varepsilon_{n\bm{k}} - E_\text{F})\delta(\varepsilon_{m\bm{k+q}} - E_\text{F})
    \,,
\end{equation}
The quantities $\lambda^{\mathrm{tr}}_{\bm{q}\nu}$ and $\lambda_{\bm{q}\nu}$ denote the mode-resolved transport and Eliashberg electron-phonon coupling constants, respectively. Here, $\varepsilon_{n\bm{k}}$ are the Kohn--Sham eigenvalues, $g_{mn,\nu}(\bm{k}, \bm{q})$ are the electron--phonon coupling matrix elements, $E_\text{F}$ is the Fermi energy, and $V_\text{BZ}$ is the volume of the Brillouin zone. The phonon frequency is denoted by $\omega_{\bm{q}\nu}$ for mode $\nu$ and wavevector $\bm{q}$. The electronic group velocities are given by $\bm{v}_{n\bm{k}} = \nabla_{\bm{k}} \varepsilon_{n\bm{k}}$.

\end{widetext}

We therefore argue that the Eliashberg $\lambda$ allows us to reliably approximate the transport relaxation time, and that incorporating it alongside $\sigma_{\mathrm{BT}}$ yields a more faithful representation of the true electrical conductivity than either descriptor alone.

While both $\sigma_{\mathrm{BT}}$ and $\lambda$ can, in principle, be evaluated from first principles, their computation across millions of compounds is prohibitively expensive. The BoltzTraP2 calculation of $\sigma_{\mathrm{BT}}$ requires the electronic band structure on a sufficiently dense $\mathbf{k}$-point grid to ensure the accuracy of the Fourier interpolation of the relevant band-structure quantities; $\lambda$, on the other hand, requires the linear response of the Kohn--Sham wavefunctions to atomic displacements within density-functional perturbation theory, which is among the most computationally demanding steps in standard ab initio workflows. We therefore develop and employ accurate and efficient machine learning surrogate models for each quantity independently.
For $\sigma_{\mathrm{BT}}$, no pre-existing model was available, and we therefore trained one from scratch. We used the approximately 500\,000 electronic structure calculations available in the Alexandria database~\cite{pbesolscandataset, alex1, alex2}, computed using the PBEsol functional~\cite{Perdew2008}. These electronic structure calculations were performed with $\mathbf{k}$-point meshes corresponding to a density of 8\,000 $\mathbf{k}$-points per reciprocal atom, which is sufficient to provide a reliable basis for BoltzTraP2 transport calculations. Of this initial set, 341\,000 compounds were identified as metallic and subsequently used as training data for the surrogate model of $\sigma_{\mathrm{BT}}$. For $\lambda$, we make use of a machine learning model previously trained on a large dataset of ab initio electron--phonon calculations. This model predicts $\lambda$ directly from the crystal structure, entirely bypassing the need for explicit lattice-dynamical calculations and making it well suited for deployment in a high-throughput screening context.

As the material pool for screening, we employed compounds from the Alexandria database whose geometries were relaxed using the Orb~\cite{orb} (version 2) universal machine learning interatomic potential (uMLIP; see \onlinecite{newalexpaper} for details). This uMLIP offers a computational efficiency several orders of magnitude greater than that of conventional DFT codes while retaining high accuracy, particularly with respect to equilibrium geometries. From this pool, we selected those with a distance to the convex hull of thermodynamic stability not exceeding 25~meV/atom, chosen to focus the screening on thermodynamically accessible phases. This selection yields a dataset of approximately 2.8 million compounds, for which both descriptors were evaluated using the respective surrogate models described above.

\begin{figure}[htbp]
\centering
\includegraphics[width=0.99\columnwidth]{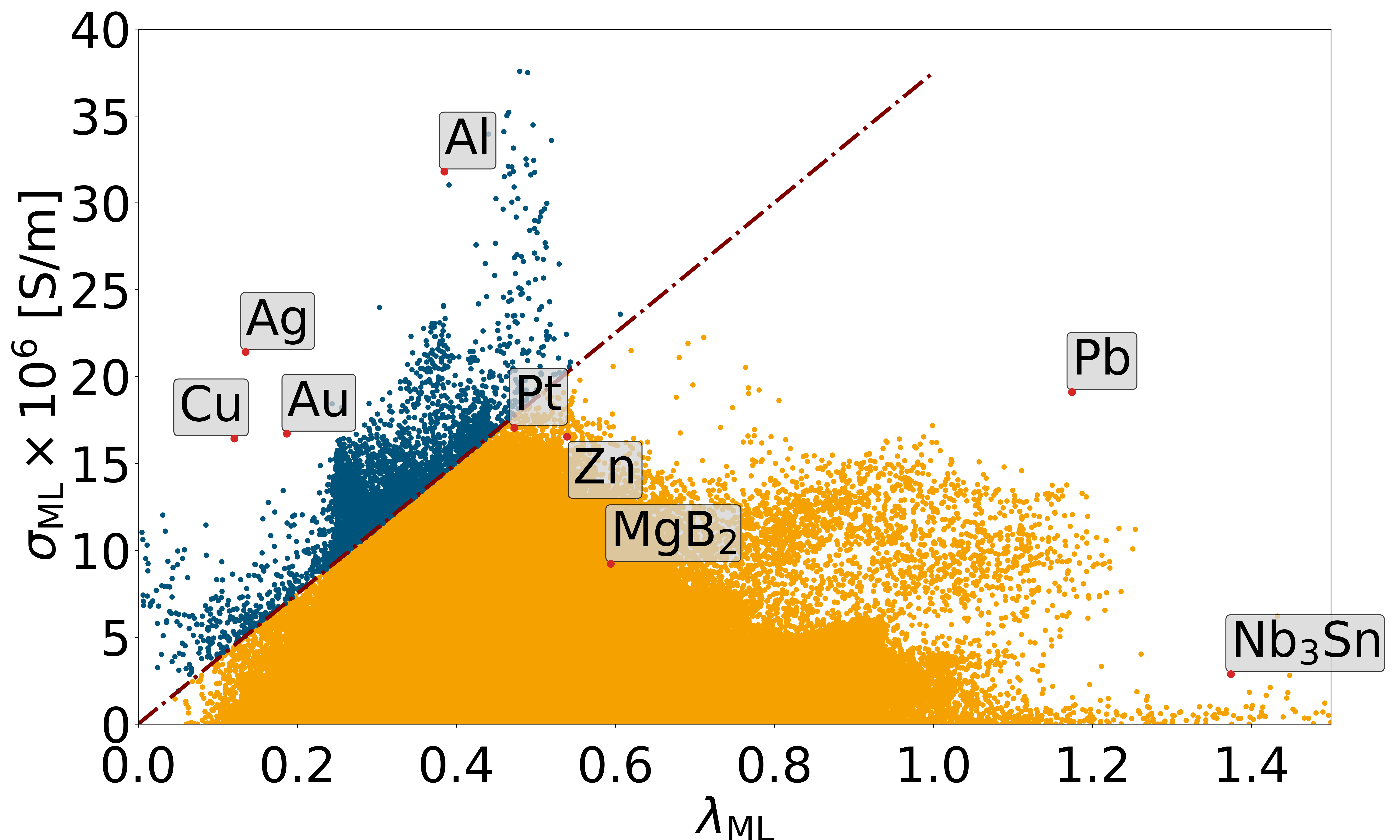}
\caption{Electronic contribution to the electrical conductivity, $\sigma^\text{ML}$, versus the superconducting electron--phonon coupling constant $\lambda^\text{ML}$ for compounds in the Alexandria database, as predicted by machine-learning models. \new{Compounds predicted by the fitted model of Fig.~\ref{fig:datafit} to exceed $\sigma_{\mathrm{SR}} = 15 \times 10^{6}\,\mathrm{S/m}$ are shown in blue.}}
\label{fig:sigma_lambda}
\end{figure}

The resulting distribution of $\sigma_\text{BT}$ and $\lambda$ obtained through the machine learning models is shown in Fig.~\ref{fig:sigma_lambda}, providing a global overview of transport behavior across a significant portion of the materials space. Ideally, the best electrical conductors should combine a large $\sigma_\text{BT}$ (large density of states at the Fermi level and large Fermi velocity) with a small $\lambda$ (low electron–phonon coupling), placing them in the top-left corner of the plot. Indeed, well-known high-conductivity metals such as Au, Ag, and Cu occupy this region, confirming the ability of our approach to correctly identify good conductors. In contrast, materials with comparatively poorer conductivity, such as Pb or \ce{Nb3Sn}, lie in the region of larger $\lambda$ and lower $\sigma_\text{BT}$. The overall distribution suggests that there are many excellent electronic conductors, but also that the likelihood of discovering materials with conductivities exceeding that of the best known conductors is low.

\begin{figure}[htbp]
\centering
\includegraphics[width=0.99\columnwidth]{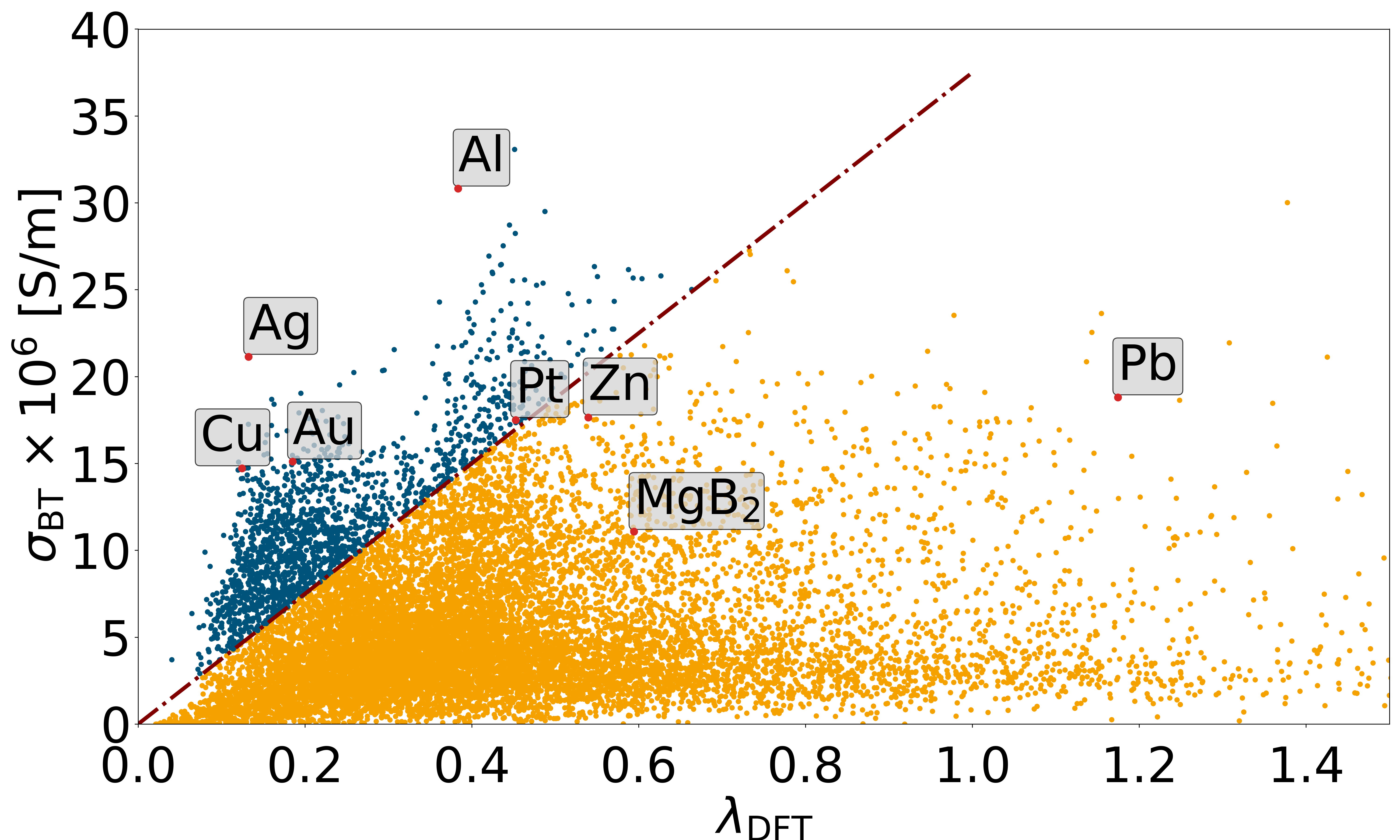}
\caption{Electronic contribution to the electrical conductivity, $\sigma_\text{BT}$, versus electron--phonon coupling $\lambda$ for $\sim$12,800 compounds from the Alexandria Database and $\sim$1,200 additional ML-selected compounds, calculated using density functional theory with the PBEsol functional. \new{Compounds predicted by the fitted model of Fig.~\ref{fig:datafit} to exceed $\sigma_{\mathrm{SR}} = 15 \times 10^{6}\,\mathrm{S/m}$ are shown in blue.}}
\label{fig:sigma_lambda_dft}
\end{figure}

We performed first-principles validation calculations for all compounds whose machine-learning-predicted conductivity exceeded the screening threshold of $\sigma_{\mathrm{SR}} > 15 \times 10^{6}$~S/m (the blue points in Fig.~\ref{fig:sigma_lambda}). For each of these compounds, we carried out full electronic structure and BoltzTraP2 calculations to obtain $\sigma_{\mathrm{BT}}$, as well as electron--phonon calculations to evaluate the Eliashberg coupling constant $\lambda$. To this set, we added all other compounds from Alexandria for which we had previously computed $\sigma_{\mathrm{BT}}$ and $\lambda$. This validation dataset comprises approximately 14\,000 compounds, and the resulting joint distribution of calculated electrical conductivities and $\lambda$ values is presented in Fig.~\ref{fig:sigma_lambda_dft}.

Overall, the first-principles validation results are in good agreement with the machine learning predictions, with the overlapping data points exhibiting root-mean-square deviations of \(4.14 \times 10^{6}\,\mathrm{S/m}\) and \(0.18\) for \(\sigma\) and \(\lambda\), respectively.
The most extreme outliers present in Fig.~\ref{fig:sigma_lambda}, and especially those for which the machine learning model predicted anomalously large values of $\sigma_{\mathrm{BT}}$, were false positives successfully identified and eliminated by the validation step. The surrogate model for $\lambda$ exhibits a higher predictive accuracy than that for $\sigma_{\mathrm{BT}}$, a discrepancy that can be attributed to the sensitivity of the BoltzTraP2 calculations to the  $\mathbf{k}$-point sampling. This can lead to inaccurate values of $\sigma_{\mathrm{BT}}$ for some compounds, which propagate into the training data and reduce the fidelity of the corresponding surrogate model. Nevertheless, the first-principles validation step effectively corrects for these inaccuracies.

\subsection*{Symbolic Regression}
\label{subsec:symbolic}
\begin{figure}[htbp]
\centering
\includegraphics[width=0.99\columnwidth]{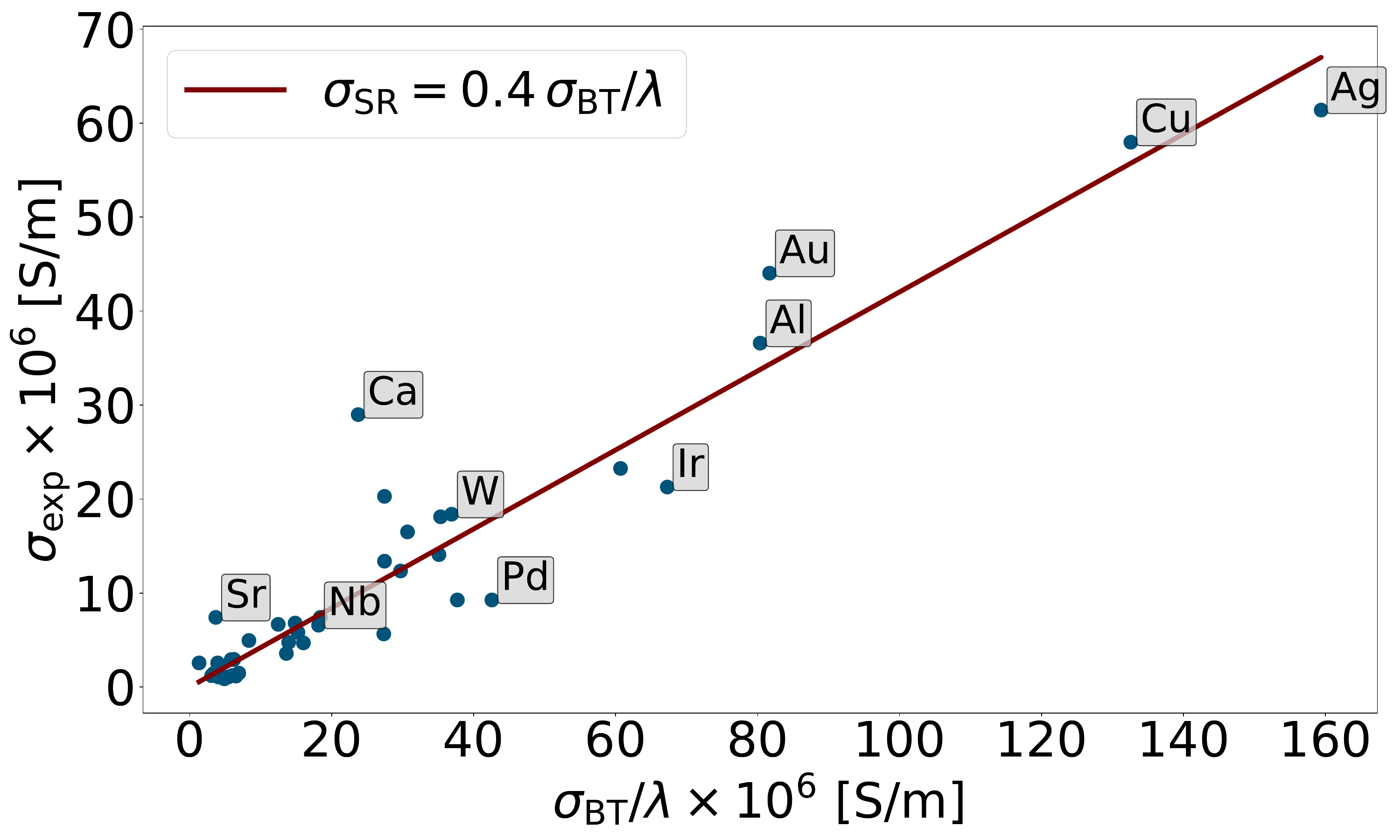}
\caption{\new{Symbolic regression (SR) of the experimental conductivity as a function of the calculated variables $\sigma_{\mathrm{BT}}$ and $\lambda$, based on a set of 43 data points.}}
\label{fig:datafit}
\end{figure}

Having evaluated both descriptors for the full pool of candidate compounds, another step in the workflow is to rank these compounds according to their estimated electrical conductivity. This requires combining $\sigma_{\mathrm{BT}}$ and $\lambda$ into a single predictive expression for $\sigma$.

We constructed a reference dataset of materials with experimental electrical conductivity data and computed $\sigma_{\mathrm{BT}}$ and the Eliashberg electron--phonon coupling constant $\lambda$ for each compound. These descriptors were used in a symbolic regression to derive an empirical relation for $\sigma_{\mathrm{exp}}$ as a function of $\sigma_{\mathrm{BT}}$ and $\lambda$, with expression complexity minimized for simplicity and interpretability. We emphasize that the symbolic regression procedure was allowed full flexibility in selecting the analytical form of the resulting model.

This procedure yields the compact relation $\sigma_\text{SR} = 0.4\,\sigma_{\mathrm{BT}}/\lambda$, in which the prefactor is dimensionless. The quality of the resulting fit is illustrated in Fig.~\ref{fig:datafit}. It is remarkable that such a simple analytical expression achieves a good description of the electrical conductivity across chemically diverse metals spanning noble metals, alkali metals, and transition metals. The most noticeable outliers (in terms of relative error) are Sr and Ca, whose anomalous behavior will be discussed in more detail below.

We point that the symbolic regression yields a formula with the same functional form as the conductivity expression when the constant relaxation time $\tau$ is replaced by its high-temperature $\lambda$-dependent scattering form. Starting from
\begin{equation}
    \sigma = \frac{\tau}{\tau^{*}} \sigma_{\mathrm{BT}},
    \label{eq:sigma_tau}
\end{equation}
and using the high-temperature relation
\begin{equation}
    \tau = \frac{\hbar}{2\pi k_\text{B} T}\frac{1}{\lambda^\text{tr}},
\end{equation}
we obtain
\begin{equation}
    \sigma = \left( \frac{\hbar}{2\pi k_\text{B} T\tau^*}\right) \frac{\sigma_{\mathrm{BT}}}{\lambda^\text{tr}}.
    \label{eq:sigma_lambda}
\end{equation}
If we assume that $\lambda=\lambda^\text{tr}$, and insert the numerical values of the constants, a temperature of $T=300$~K and using BoltzTraP’s default relaxation time $\tau^{*} = 10^{-14}\,\mathrm{s}$ \cite{boltztrap2}, we obtain the prefactor $0.405$, remarkably close to the one obtained through symbolic regression. This suggests that, for the chemically diverse set of metals considered here, the difference between $\lambda$ and $\lambda_{\text{tr}}$ has a limited impact on the ranking of conductivities.

On the basis of this fit, we define a conductivity threshold of $\sigma_{\mathrm{SR}} > 15 \times 10^{6}$~S/m as the criterion for identifying candidate high-conductivity compounds in the subsequent screening. This threshold is indicated in Figs.~\ref{fig:sigma_lambda}, \ref{fig:sigma_lambda_dft}, providing a natural demarcation between excellent metallic conductors and the broader population of metals with more moderate transport properties.

\subsection*{Electrical conductivity via ab initio Boltzmann transport equation}

Having established a reliable criterion for identifying candidate high-conductivity compounds, we proceed to evaluate the electrical conductivity from first principles for the most promising materials. These calculations are performed using the \textsc{EPW} code~\cite{epw1,epw2}, which implements the linearized BTE to describe the first-order response of the electronic distribution function to an applied electric field. The required inputs are the electronic band structure and the electron--phonon coupling matrix elements as a function of crystal momentum and band index, both of which are obtained from DFT calculations. Since ab initio calculations of the electrical conductivity remain scarce in the literature, we first carry out a series of benchmark calculations to assess the accuracy of this approach before applying it to novel compounds.

Despite electrical conductivity being among the longest-studied transport properties in condensed matter physics, reliable experimental reference data for stoichiometric, well-ordered metallic compounds remain surprisingly limited. We therefore benchmark our calculations against a carefully selected set of elemental metals for which well-established experimental conductivity values are available~\cite{lide2005crc}. The calculated room-temperature conductivities are summarized in Table~\ref{tab:benchmark} and compared against the corresponding experimental references.

\begin{table}[htbp]
\centering
\caption{Calculated and experimental~\cite{lide2005crc} electrical conductivity $\sigma$ $(10^{6}\,\mathrm{S/m})$ at ambient temperature (300~K), with a mean absolute error (MAE) of $2.97 \times 10^{6}\,\mathrm{S/m}$ between $\sigma_\text{EPW}$ and $\sigma_\text{exp}$. }
\label{tab:benchmark}
\begin{tabular*}{.95\columnwidth}{@{\extracolsep{\fill} }lrr@{\hspace{.6cm}}c@{}lrr}
System & $\sigma_\text{EPW}$ & $\sigma_\text{exp}$ & &System & $\sigma_\text{EPW}$ & $\sigma_\text{exp}$\\
\toprule
Cu& 60.84  & 57.97 &   & Pt& 12.10  & 9.26    \\
Ag& 52.06  & 61.39 &   & Th&  9.87  & 6.80    \\
Au& 41.89  & 44.03 &   & Pb&  8.74  & 4.69    \\ 
Al& 39.47  & 36.59 &   & Ca&  7.98  & 28.99   \\
Na& 26.16  & 20.28 &   & Ta&  7.36  & 7.41    \\
Rh& 24.39  & 23.26 &   & Nb&  6.62  & 6.58    \\
Mg& 22.80  & 22.17 &   & V &  3.28  & 4.95    \\
Ir& 20.82  & 21.27 &   & Ba&  3.07  & 2.92    \\
Mo& 18.42  & 18.12 &   & Cs&  2.88  & 4.76    \\
W & 16.75  & 18.38 &   & Sr&  2.65  & 7.41    \\
K & 13.18  & 13.39 &   & Ti&  1.91  & 2.56    \\
Pd& 12.39  & 9.26  &   & Ce&  1.80  & 1.21--1.34     \\
\bottomrule
\end{tabular*}
\end{table}

The results reproduce the experimental reference values with good overall accuracy, with most computed values falling within 30\% of the measured data. The agreement is relatively strong for metals such as Cu, Au, and Mo, where the calculated and experimental conductivities are in close quantitative agreement.

\new{The largest deviations are observed for the alkaline-earth metals Ca and Sr, which also appeared as outliers in Fig.~\ref{fig:datafit}. A plausible origin lies in the description of the underlying band structure. Although both elements are commonly classified as $sp$ metals, their empty $d$ states lie only a few\,eV above $E_{\mathrm{F}}$ and hybridise with the conduction band. Quasiparticle corrections of several hundred meV in the vicinity of $E_{\mathrm{F}}$ have been reported for both metals~\cite{Friedrich2022}, and since their Fermi surfaces intersect the Brillouin-zone boundaries, shifts of this magnitude can alter the Fermi-surface topology and, with it, the Fermi velocities and the phase space available for electron--phonon scattering.}

\begin{figure}[htbp]
    \centering
    \includegraphics[width=0.99\linewidth]{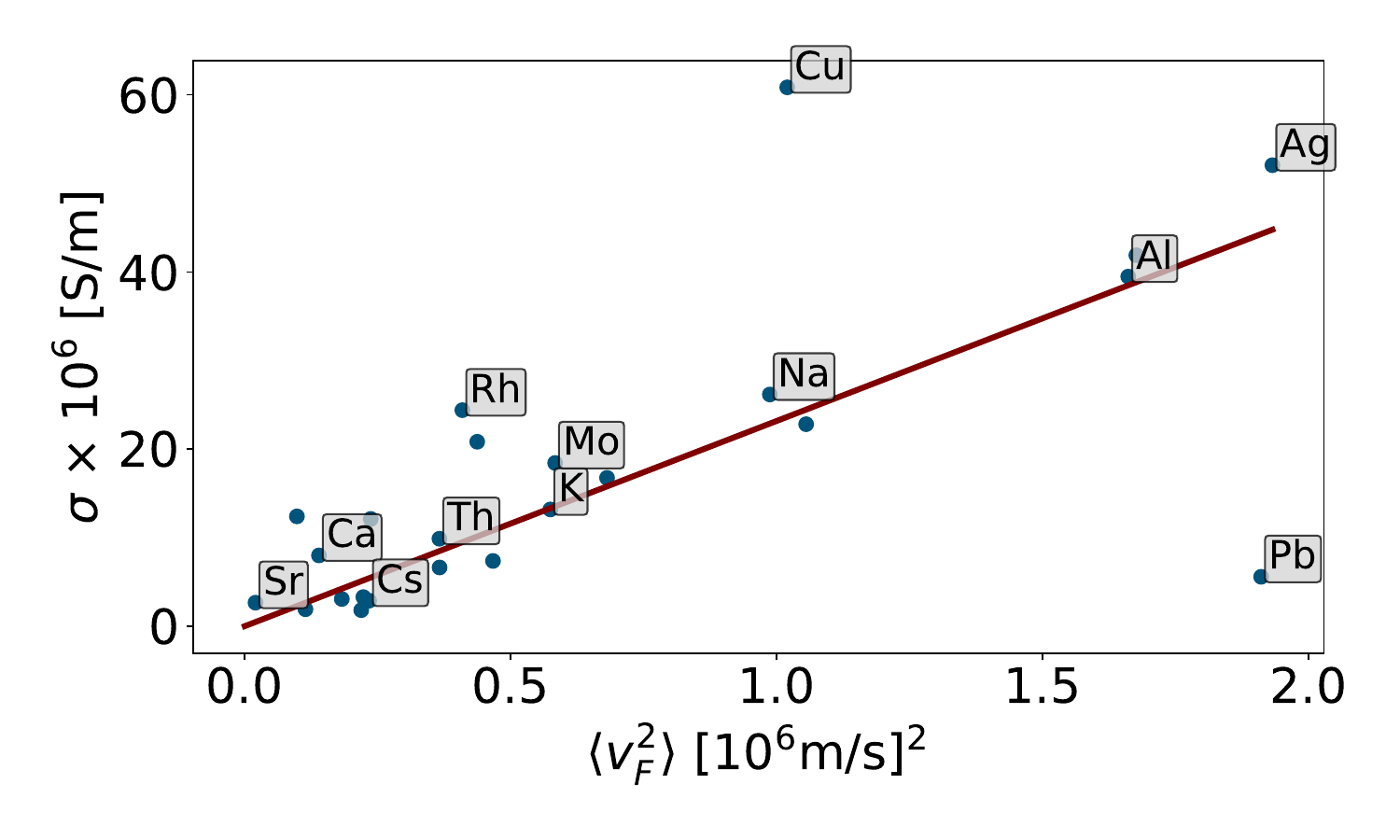}
    \caption{\new{Electrical conductivity $\sigma$ of the elemental metals as a function of the average squared Fermi velocity $\langle v_F^{2} \rangle$.}}
    \label{fig:fvel}
\end{figure}

\new{Besides the agreement with experiment, it is interesting to examine how the computed conductivities relate to the Fermi velocity. Figure~\ref{fig:fvel} shows that, for these simple isotropic metals, the electrical conductivity $\sigma$ obtained from EPW transport calculations tends to scale with the Fermi-surface average of the squared Fermi velocity, $\langle v_F^2 \rangle$, in line with the semiclassical Boltzmann transport picture, in which more energetic carriers propagate charge more efficiently between scattering events. Cu and Pb are the clearest exceptions to this trend, lying well above and well below, respectively, the line defined by the remaining metals. This mirrors the electron--phonon coupling strengths reported earlier (Fig.~\ref{fig:sigma_lambda}) as Cu's weak coupling pushes $\sigma$ above the $\langle v_F^2 \rangle$ prediction, while Pb's strong coupling pushes it below.}

\renewcommand{\arraystretch}{1.02}
\begin{table}[htbp]
\centering
\caption{Electrical conductivity calculated with EPW $\sigma_\text{EPW}$ $(10^{6}\,\mathrm{S/m})$ and fitted model estimated $\sigma_{\mathrm{SR}}$ = $0.4\,\sigma_{BT}/\lambda$ at 300~K. The MAE between $\sigma_\text{EPW}$ and $\sigma_{\mathrm{SR}}$ is $3.82 \times 10^{6}\,\mathrm{S/m}$.}
\label{tab:conductivity_300K}
\begin{tabular*}{.95\columnwidth}{@{\extracolsep{\fill}} lrr@{\hspace{.6cm}}c@{}lrr}
System & $\sigma_\text{EPW}$ & $\sigma_{\mathrm{SR}}$  & &System & $\sigma_\text{EPW}$ & $\sigma_{\mathrm{SR}}$ \\
\toprule
\ce{Be2IrPt}   & 38.02 & 30.32 &  & \ce{BeNi2Zn}   & 22.56 & 22.11 \\
\ce{H3NaNi}    & 34.82 & 25.74 &  & \ce{V2Re}      & 22.33 & 16.04 \\
\ce{LuIr}      & 34.41 & 39.56 &  & \ce{NCaNi}     & 21.44 & 17.64 \\
\ce{LiBePt2}   & 34.35 & 33.17 &  & \ce{Co2ZnGa}   & 21.26 & 15.93 \\
\ce{CuZnPd2}   & 33.93 & 32.09 &  & \ce{MgAl3}     & 20.95 & 22.96 \\
\ce{LiAlRhPt}  & 33.22 & 31.29 &  & \ce{NbMo}      & 20.91 & 18.25 \\
\ce{MgAlRh2}   & 33.15 & 29.13 &  & \ce{Nb2Re}     & 20.47 & 16.71 \\
\ce{Be2RhPt}   & 31.57 & 29.58 &  & \ce{LiNiZn2}   & 20.37 & 19.16 \\
\ce{LiMgPd2}   & 30.13 & 31.04 &  & \ce{NbTaWRe}   & 20.35 & 15.76 \\
\ce{H3CaNi}    & 29.49 & 17.70 &  & \ce{PCo2}      & 20.02 & 17.18 \\
\ce{ErIr}      & 29.28 & 35.07 &  & \ce{LiSiCoNi}  & 19.59 & 17.18 \\
\ce{ZnGaRh2}   & 28.90 & 32.02 &  & \ce{LiNi2Ga}   & 19.42 & 19.77 \\
\ce{AlZnRh2}   & 28.82 & 30.34 &  & \ce{V2WRe}     & 19.38 & 16.63 \\
\ce{MgPd2Cd}   & 28.15 & 32.58 &  & \ce{MgIn}      & 19.33 & 16.25 \\
\ce{ZnPt}      & 27.79 & 30.44 &  & \ce{LiZn2Cu}   & 19.15 & 25.60 \\
\ce{ZnRh2In}   & 27.39 & 29.22 &  & \ce{O3Re}      & 18.18 & 20.99 \\
\ce{Mg2RhPt}   & 27.38 & 29.41 &  & \ce{Mg2CdIn}   & 17.74 & 16.30 \\
\ce{NbW2}      & 26.72 & 19.85 &  & \ce{CoNiZn2}   & 17.41 & 14.99 \\
\ce{AlCo}      & 26.30 & 20.43 &  & \ce{LiAl3}     & 17.19 & 17.94 \\
\ce{H3NaRh}    & 26.28 & 39.89 &  & \ce{LiMg2}     & 16.04 & 15.96 \\
\ce{PdCd}      & 25.37 & 29.48 &  & \ce{LiSiNi2}   & 15.53 & 16.77 \\
\ce{MgAg}      & 25.25 & 30.45 &  & \ce{MgHg}      & 15.49 & 17.70 \\
\ce{Be2CoNi}   & 25.07 & 21.42 &  & \ce{Li2NiSn}   & 14.48 & 14.85 \\
\ce{B2Al}      & 23.79 & 15.72 &  & \ce{AlSiNi}    & 14.22 & 15.22 \\
\ce{MgPd}      & 22.63 & 30.90 &  & \ce{BeTiGa2}   & 13.96 & 15.05 \\
\ce{MoTa}      & 22.61 & 18.71 &  & \ce{BeNi}      & 13.69 & 24.68 \\
\bottomrule
\end{tabular*}
\end{table}

We now turn to the most promising compounds identified in the Symbolic Regression section above. To ensure broad coverage of chemical space while avoiding over-representation of closely related compositions, we retained a single representative for each group of compounds with identical elemental composition. Specifically, we selected the compound with the highest predicted conductivity within each system. Full ab initio iBTE calculations were subsequently performed on this selected set using \textsc{EPW}.

The results of the calculations are summarized in Table~\ref{tab:conductivity_300K}. The calculated values confirm that the selected materials exhibit high electrical conductivity, with several candidates showing predicted conductivities comparable to that of aluminum ($36.59 \times 10^{6} \mathrm{S/m}$) at room temperature.

Although none of the identified materials achieves conductivities comparable to those of the best known metals (Ag, Au, and Cu), the top candidates, \ce{Be2IrPt}, \ce{H3NaNi}, \ce{LuIr}, and \ce{LiBePt2} reach conductivities ranging from $\sim 34$ to $38 \times 10^{6}\,\mathrm{S/m}$, placing them in the range of highly conductive compounds.

\new{All results reported above were obtained at the scalar-relativistic level, as spin--orbit coupling (SOC) was neglected. Since the SOC strength grows rapidly with nuclear charge, it is expected to be non-negligible for the heavy elements (Ir, Pt, Au, Hg, Pb) occurring in several of the compounds considered here. To quantify this approximation, the transport calculations were repeated with SOC included for \ce{Pb}, and for \ce{Be2IrPt} and \ce{LiBePt2}, two of the highest-conductivity systems in Table~\ref{tab:conductivity_300K}, each containing a heavy $5d$ element. For \ce{Pb}, the room-temperature conductivity obtained with SOC is $4.04 \times 10^{6}\,\mathrm{S/m}$, which is in much better agreement with the experimental result. For the two compounds, the room-temperature conductivities are reduced from $38.02$ and $34.35 \times 10^{6}\,\mathrm{S/m}$ to $34.96$ and $31.14 \times 10^{6}\,\mathrm{S/m}$, corresponding to decreases of $8.1\%$ and $9.3\%$, respectively. Although relativistic effects have a moderate influence on the transport properties of heavy-element compounds, given the substantially higher computational cost of electron--phonon calculations including SOC, its neglect thus represents an acceptable trade-off within the high-throughput workflow adopted here.}
\begin{figure}[tbh]
\centering
\includegraphics[width=0.45\textwidth]{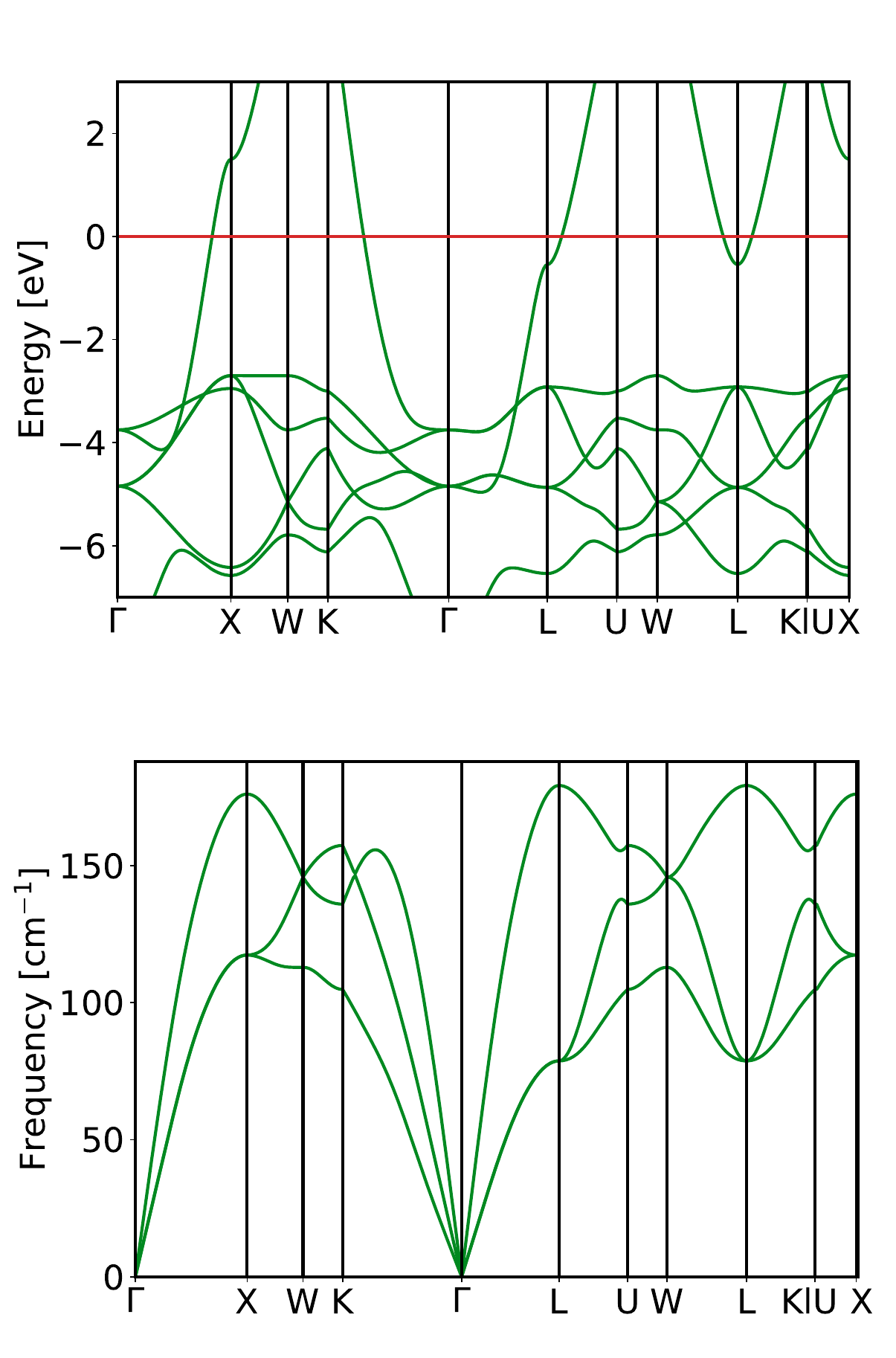}
\caption{Electronic (top) and phonon (bottom) band structures of Ag.}
\label{fig:bs_ag}
\end{figure}

\begin{figure}[tbh]
\includegraphics[width=0.45\textwidth]{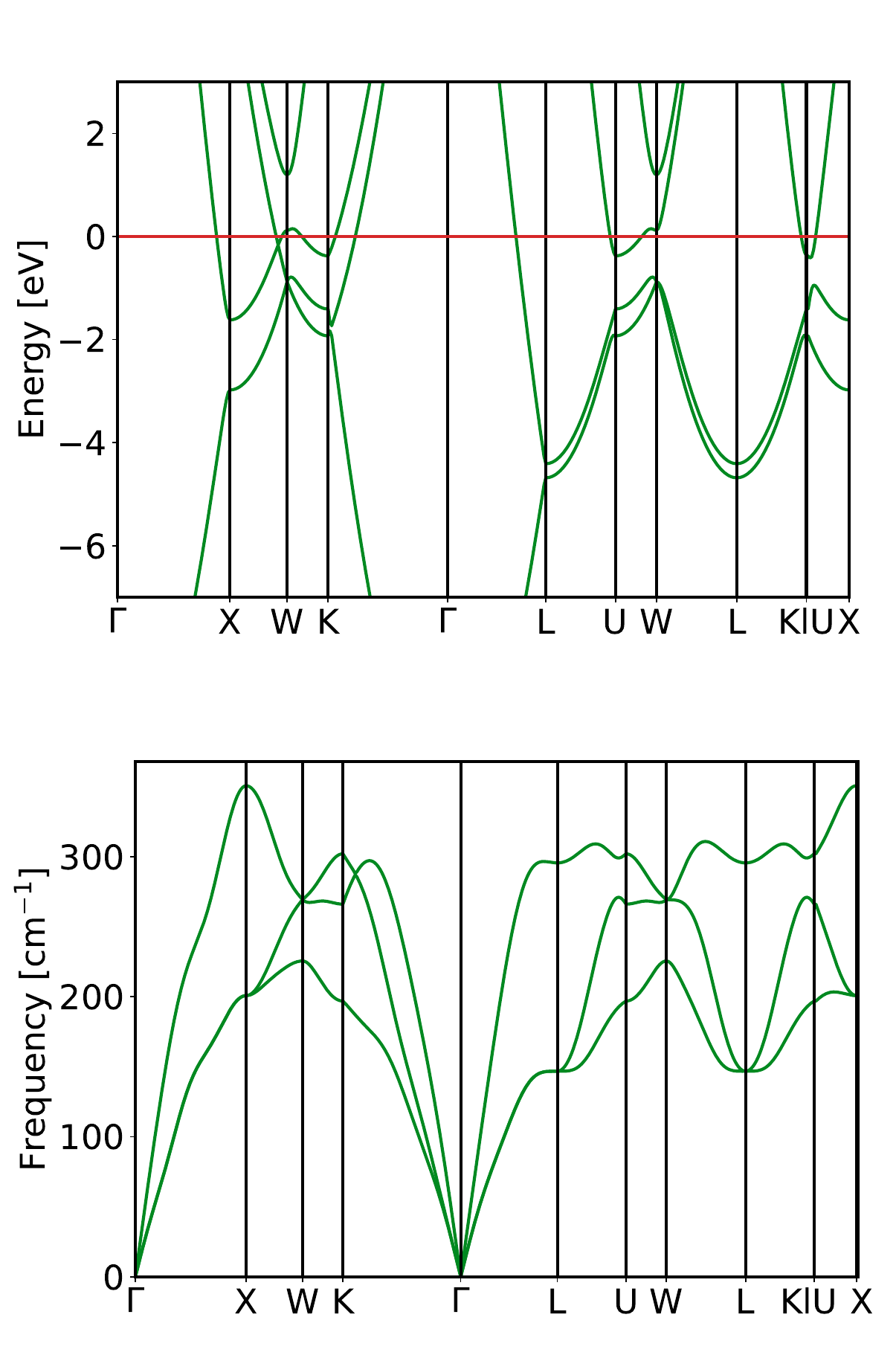}
\caption{Electronic (top) and phonon (bottom) band structures of Al.}
\label{fig:bs_al}
\end{figure}

Silver exhibits the highest electrical conductivity among elemental metals at ambient conditions, and our screening correctly flags it among the top-ranked conductors. This exceptional behavior originates from the combination of its electronic structure and lattice dynamics (see Fig.~\ref{fig:bs_ag}). Electronically, silver possesses a closed $4d^{10}$ shell, resulting in a complex of five narrow $d$-bands situated approximately 4--7~eV below the Fermi level ($\varepsilon_F$). Consequently, the states at $\varepsilon_F$ are dominated by the single $5s$ electron, which forms a broad, parabolic, nearly-free-electron band. The steep dispersion of this band yields an exceptionally high Fermi velocity ($v_F$), providing a high intrinsic capacity for charge transport. The phonon band structure (see Fig. \ref{fig:bs_ag}) shows relatively low vibrational frequencies due to the large atomic mass of Ag, with a Brillouin zone boundary maximum below 150~cm$^{-1}$. Crucially, these phonons couple very weakly to the conduction electrons, resulting in a remarkably small electron--phonon coupling constant ($\lambda \approx 0.1$). This weak coupling facilitates a long electronic relaxation time ($\tau$), which, when paired with the high $v_F$, is responsible for the exceptional electrical conductivity of silver.

It is instructive to compare Ag (see Fig. \ref{fig:bs_ag}) and Al (see Fig. \ref{fig:bs_al}), as they represent two distinct paradigms in metallic conduction. The Fermi surface of Al is significantly more complex than that of Ag, consisting of multiple sheets, which might be indicative of stronger electronic scattering channels. Al has a much stronger electron--phonon coupling constant ($\lambda \approx 0.45$) compared to Ag. A manifestation of this interaction is that Al becomes a superconductor at a transition temperature of $1.2$~K, whereas silver does not exhibit superconductivity. This same coupling that facilitates Cooper pair formation at cryogenic temperatures manifests as a higher scattering rate at ambient conditions, thereby limiting the room-temperature electrical conductivity of Al relative to Ag.

\begin{figure}[tbh]
\centering
\includegraphics[width=0.45\textwidth]{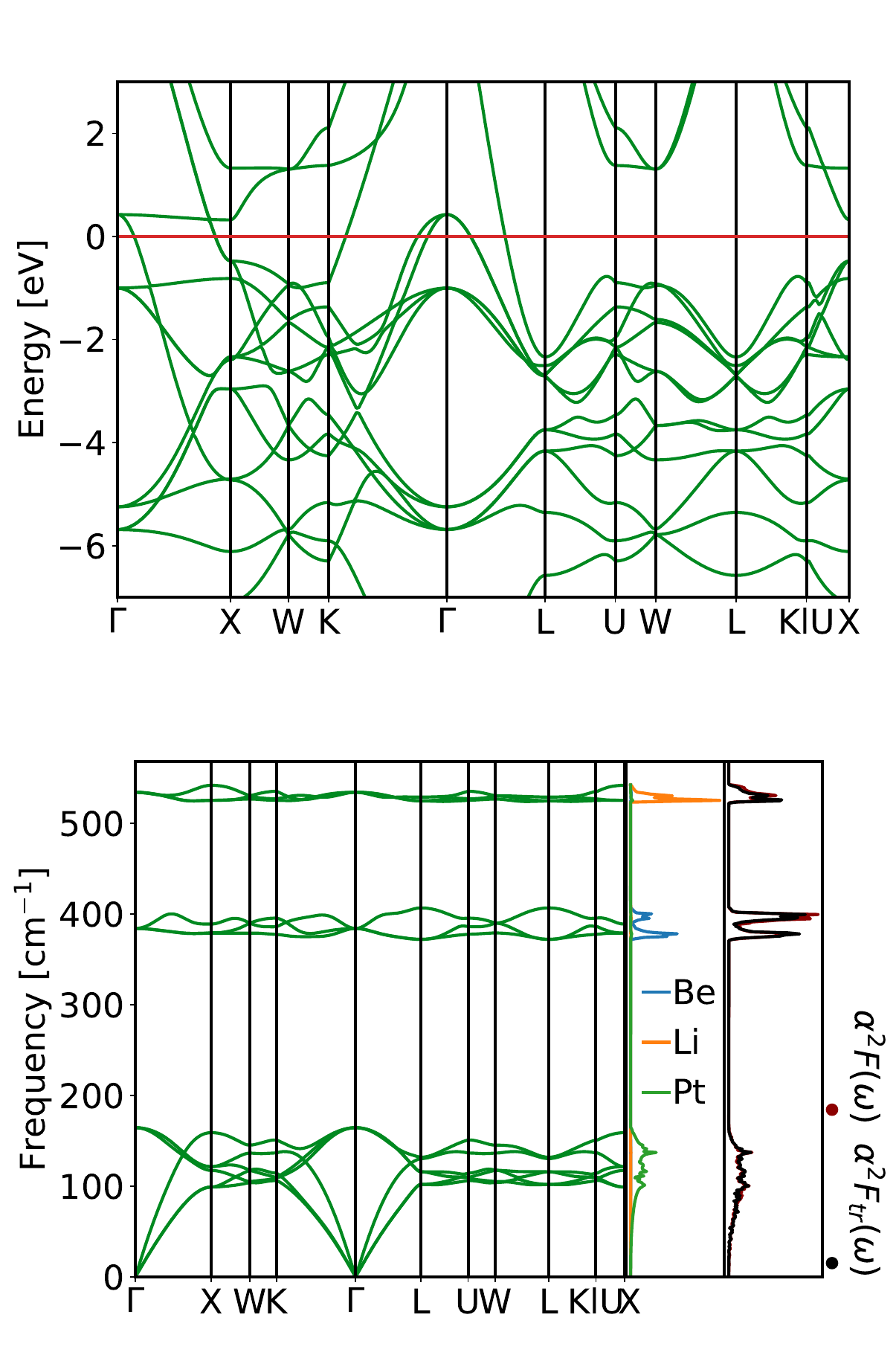}
\caption{Electronic (top) and phonon (bottom) band structures of LiBePt$_2$.}
\label{fig:bs_libept2}
\end{figure}

\begin{figure}[tbh]
\includegraphics[width=0.45\textwidth]{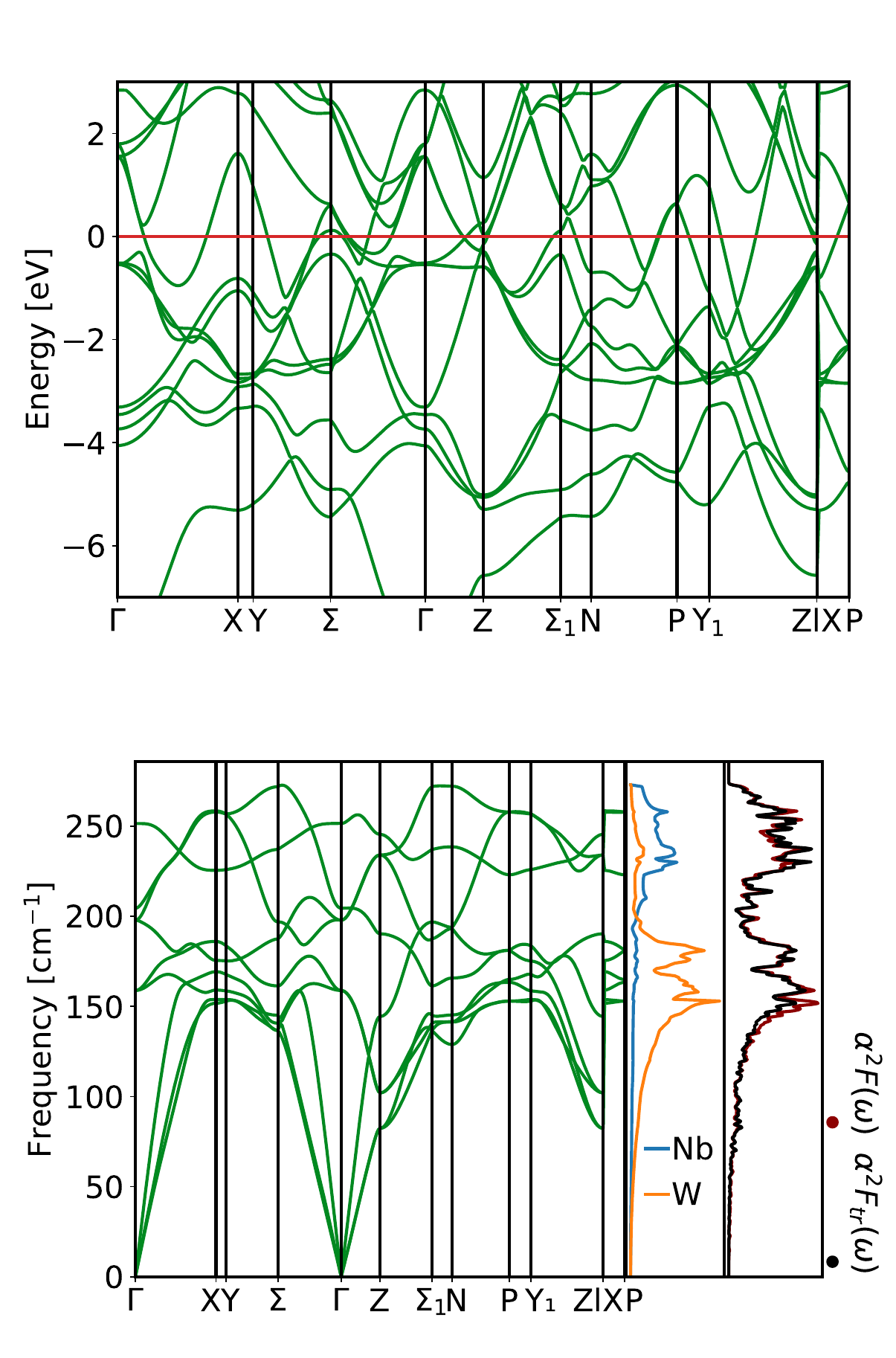}%
\caption{Electronic (top) and phonon (bottom) band structures of NbW$_2$.}
\label{fig:bs_nbw2}
\end{figure}

\ce{LiBePt2} stands out among the top-performing compounds in our ranking, with a predicted electrical conductivity of $34.35 \times 10^6$~S/m, and it is also one of the cases where the symbolic-regression model closely reproduces the EPW reference ($33.17 \times 10^6$~S/m). The underlying electronic and phononic mechanisms for this behavior are illustrated in Fig.~\ref{fig:bs_libept2}. Electronically, the states near the Fermi level are dominated by Pt orbitals. However, the additional valence electrons provided by Li and Be shift the Fermi level upward, effectively pushing the high-density $d$-states beneath $\varepsilon_F$.

Three distinct bands cross $\varepsilon_F$: a highly dispersive band and two bands that are degenerate at the $\Gamma$ point and that form a carrier pocket around the zone center. The phonon band structure is characterized by three distinct manifolds, well-separated due to the large disparities in atomic mass. The low-frequency region ($<200$~cm$^{-1}$) is dominated by heavy Pt atoms, while the mid-frequency ($\sim 400$~cm$^{-1}$) and high-frequency ($>500$~cm$^{-1}$) regions are attributed to Be and the light Li atoms, respectively. The Eliashberg spectral function $\alpha^2F(\omega)$ reveals that scattering is primarily concentrated within the low-frequency Pt-dominated modes. This concentration yields a relatively small electron–phonon coupling constant ($\lambda \approx 0.156$), which indicates that vibrations of the Pt sublattice serve as the principal source of resistivity. 

The compound $\text{NbW}_2$ is of particular interest as its predicted electrical conductivity of $26.72 \times 10^6$~S/m exceeds those of its individual constituents. While the predicted conductivities for pure Nb and W are $6.62 \times 10^6$~S/m (experimental: $6.58 \times 10^6$~S/m) and $16.75 \times 10^6$~S/m (experimental: $18.38 \times 10^6$~S/m), respectively, $\text{NbW}_2$ demonstrates markedly superior transport properties. The electronic and phononic band structures are illustrated in Fig.~\ref{fig:bs_nbw2}.

Multiple dispersive bands cross $\varepsilon_F$, ensuring a high concentration of itinerant charge carriers. The phonon band structure reveals a continuous spectrum of vibrational modes extending to approximately 275~cm$^{-1}$. The Eliashberg spectral function $\alpha^2F(\omega)$ indicates that electron--phonon scattering is primarily concentrated in the mid-to-high frequency range (150--250~cm$^{-1}$), with significant contributions from both Nb and W vibrational modes. This interaction results in a relatively small integrated electron--phonon coupling constant of $\lambda \approx 0.163$.

\section*{Discussion}

The results presented in this study demonstrate that modern \textit{ab initio} methods, specifically when coupled with the semi-classical Boltzmann transport theory as implemented in BoltzTraP2, \textsc{EPW} and density-functional perturbation theory for the electron-phonon interaction, provide a good description of metallic transport. The agreement between our calculated values and the experimental benchmarks across a chemically diverse set of elemental metals shows the predictive power of this approach. This quantitative alignment suggests that such first-principles methods can be reliably used for the \textit{a priori} prediction of transport properties in novel material phases where experimental data is currently unavailable.

A significant finding of our analysis is the surprising robustness of the Constant Relaxation Time Approximation (CRTA). While the CRTA omits the energy and momentum dependence of scattering, our results suggest that for a wide range of intermetallic compounds, the variation in the electronic density of states and Fermi velocity plays a more dominant role in determining the ranking of conductivity than the fine details of the scattering. By utilizing this approximation within our machine learning (ML) pipeline, we were able to screen 2.8 million compounds. The success of these ML models in identifying high-conductivity candidates confirms that descriptors derived from the electronic structure are sufficient to navigate the vast chemical space of potential conductors.

\new{However, our study also reinforces the exceptional standing status of the noble metals. Silver, alongside gold and copper, sits among the highest electrical conductivities found in our calculations. This is due to their unique electronic topology, exhibiting a single, highly dispersive $s$-band crossing the Fermi level and deeply buried $d$-states that minimize interband scattering. Because these properties are intrinsic to the atomic and crystalline symmetry of the elements, any processing, alloying, or microstructuring is inherently likely to introduce scattering centers or modify the band dispersion in a way that reduces conductivity. This suggests that Ag, Cu, and Au may be difficult to surpass for metallic conduction at ambient conditions.}

\section*{Methods}
\label{sec:methods}

\subsection*{Ab-initio calculations}

The starting point for the BoltzTraP2 calculations is given by PBEsol electronic structure calculations performed using the computational settings reported in Ref.~\cite{pbesolscandataset}, which are detailed as follows. DFT calculations were carried out within the projector augmented wave (PAW) method, as implemented in the Vienna Ab initio Simulation Package (VASP). The choice of pseudopotentials followed Ref~\cite{pbesolscandataset}. All calculations included spin polarization and were initialized in a ferromagnetic configuration. Brillouin zone integrations were performed using first-order Methfessel–Paxton smearing with a width of 0.2 eV. Total energies were evaluated using a plane-wave cutoff energy of 520 eV and dense $\Gamma$-centered k-point meshes corresponding to approximately 8000 k-points per reciprocal atom. 

Phonon and electron–phonon coupling calculations were performed using density functional theory as implemented in the \textsc{Quantum ESPRESSO} package (version 7.1)~\cite{Giannozzi2009,Giannozzi2017}. The exchange–correlation functional was treated within the PBEsol generalized gradient approximation~\cite{Perdew2008}, and norm-conserving scalar-relativistic pseudopotentials from the \textsc{PseudoDojo} project~\cite{vanSetten2018pseudodojo} (stringent set) were employed. \new{For the SOC tests the norm-conserving fully-relativistic stringent pseudopotentials were used instead, also from the \textsc{PseudoDojo} project.} The plane-wave kinetic energy cutoff was determined from the PseudoDojo recommendations for the stringent pseudopotentials. Brillouin zone integrations were carried out using Methfessel–Paxton smearing with a width of 0.2 eV. The electronic $\mathbf{k}$-point grids were defined with a density on the order of $3 \times (6\pi)^2$ KPPA, while the phonon $\mathbf{q}$-point grids were taken as half of the corresponding $\mathbf{k}$-point density. Where necessary, denser grids were used until the Wannier-interpolated band structures showed good agreement with the directly calculated ones. The converged grid for each structure is reported in the Supplementary Information.

\new{Following the Wannierization procedure~\cite{wannier90}, the BTE and electron–phonon interactions were evaluated using the \textsc{EPW} code~\cite{epw1}. The electron–phonon matrix elements were interpolated onto dense grids where the fine $\mathbf{k}$- and $\mathbf{q}$-point meshes were chosen to have 4 times as many points along each direction. Electronic occupations were treated using the adaptive smearing scheme of EPW following Eq. (18) of Ref. \cite{Li2014}, while phonon integrations over the Brillouin zone employed a Gaussian smearing of 0.05 meV. Convergence and quality tests can be found in the Supplementary Information.}

\subsection*{ALIGNN model}

\new{We used two separate \textsc{ALIGNN}~\cite{alignn} models: one, trained within this work, to predict the electrical conductivity, and another, previously trained in the context of conventional superconductivity, to predict the electron-phonon coupling parameter $\lambda$~(see \cite{Pires2026} for more detail).} For the electrical conductivity, materials were selected from the PBEsol Alexandria database with a band gap smaller than than 0.1~eV. The dataset consisted of 341k points, randomly split into 90\% / 5\% / 5\% for training, validation, and testing. The target was taken as the logarithm of the electrical conductivity in units of $10^6$~S/m, and the best model was chosen based on the validation set error. The mean absolute error (MAE) on the test set of 17\,000 samples was 0.65, compared to an average value of 3.68.

For predicting the electron-phonon coupling parameter $\lambda$, the model was trained with $\lambda$, $\omega_{\rm log}$, and $T_{\rm c}$ as simultaneous targets, weighting the error for each property equally. The dataset was split randomly into 90\% / 5\% / 5\% for training, validation, and testing. The MAE on the test set was 0.151, 31.9~K, and 2.51~K for $\lambda$, $\omega_{\rm log}$, and $T_{\rm c}$, respectively.

\section*{Data availability}
\label{sec:data_ava}
The data can be accessed and/or downloaded from \url{https://alexandria.icams.rub.de/} under the terms of the \href{https://creativecommons.org/licenses/by/4.0/}{Creative Commons Attribution 4.0 License}. \new{All input and output files are available in the Materials Cloud Archive, doi:10.24435/materialscloud:4d-hv.}

\section*{Code availability}
\label{sec:code_ava}
All code and models developed in this work will be freely available upon acceptance at  \url{https://github.com/hyllios/utils/tree/main/}.

\section*{Acknowledgements}
T.H.B.S. and M.A.L.M. were supported by a collaboration between the Kavli Foundation, Klaus Tschira Stiftung, and Kevin Wells, as part of the SuperC collaboration, and by the Simons Foundation through the Collaboration on New Frontiers in Superconductivity (Grant No. SFI-MPS-NFS-00006741-10). S.B. acknowledge funding from the Volkswagen Stiftung (Momentum) through the project ‘‘dandelion''. The authors gratefully acknowledge the computing time made available to them on the high-performance computers Noctua and Otus at the NHR Center Paderborn Center for Parallel Computing (PC2). The authors gratefully acknowledge the scientific support and HPC resources provided by the Erlangen National High Performance Computing Center (NHR@FAU) of the Friedrich-Alexander-Universität Erlangen-Nürnberg (FAU) under the NHR project k114eb. Part of the calculations were performed on the HPC cluster Elysium of the Ruhr University Bochum, subsidised by the DFG (INST 213/1055-1).

\section*{Author  Contributions}
T.H.B.S., T.F.T.C., and H.-C.W. developed the workflow and performed the EPW calculations. T.F.T.C. and M.A.L.M. trained the machine learning models and carried out the predictions, while S.B. and M.A.L.M. supervised the research. T.H.B.S., T.F.T.C., H.-C.W., M.A.L.M, S.B and S.D.C contributed to the interpretation of the results and the writing of the manuscript. All authors have read and approved the manuscript.

\section*{Competing  Interests}
The authors declare that they have no competing interests.

\clearpage
\onecolumngrid
\begin{center}
{\large\bfseries References}
\end{center}
\vspace{-1cm}
\twocolumngrid
\bibliographystyle{plain}
\bibliography{bib}

\clearpage
\onecolumngrid
\begin{center}
{\large\bfseries Figure Legends}
\end{center}
\vspace{-0.2cm}

\noindent\textbf{Figure~\ref{fig:flowchart}.} \new{Schematic overview of the computational workflow. Starting from the ${\sim}2.8$~M compounds in the Alexandria database with $E_{\mathrm{hull}} < 25$~meV/atom, two ML models (a pre-existing one for the electron-phonon coupling $\lambda$ and one trained here for the BoltzTraP conductivity $\sigma_{\mathrm{BT}}$) predict $\lambda$ and $\sigma_{\mathrm{BT}}$ for the full set. An empirical fit, $\sigma_{\mathrm{SR}} = 0.4\,\sigma_{\mathrm{BT}}/\lambda$, calibrated against experimental conductivities, then estimates $\sigma_{\mathrm{exp}}$, and compounds exceeding $15\times10^{6}$~S/m are retained (${\sim}1200$~metals). These undergo explicit BoltzTraP and electron-phonon calculations and are merged with ${\sim}12{,}800$ compounds already available in the database (${\sim}14{,}000$ total). Reapplying the same fit narrows this set to 52 unique systems, for which accurate transport properties are finally computed with EPW by solving the iterative Boltzmann transport equation.}

\medskip
\noindent\textbf{Figure~\ref{fig:sigma_lambda}.} Electronic contribution to the electrical conductivity, $\sigma^\text{ML}$, versus the superconducting electron--phonon coupling constant $\lambda^\text{ML}$ for compounds in the Alexandria database, as predicted by machine-learning models. \new{Compounds predicted by the fitted model of Fig.~\ref{fig:datafit} to exceed $\sigma_{\mathrm{SR}} = 15 \times 10^{6}\,\mathrm{S/m}$ are shown in blue.}

\medskip
\noindent\textbf{Figure~\ref{fig:sigma_lambda_dft}.} Electronic contribution to the electrical conductivity, $\sigma_\text{BT}$, versus electron--phonon coupling $\lambda$ for $\sim$12,800 compounds from the Alexandria Database and $\sim$1,200 additional ML-selected compounds, calculated using density functional theory with the PBEsol functional. \new{Compounds predicted by the fitted model of Fig.~\ref{fig:datafit} to exceed $\sigma_{\mathrm{SR}} = 15 \times 10^{6}\,\mathrm{S/m}$ are shown in blue.}

\medskip
\noindent\textbf{Figure~\ref{fig:datafit}.} \new{Symbolic regression (SR) of the experimental conductivity as a function of the calculated variables $\sigma_{\mathrm{BT}}$ and $\lambda$, based on a set of 43 data points.}

\medskip
\noindent\textbf{Figure~\ref{fig:fvel}.} \new{Electrical conductivity $\sigma$ of the elemental metals as a function of the average squared Fermi velocity $\langle v_F^{2} \rangle$.}

\medskip
\noindent\textbf{Figure~\ref{fig:bs_ag}.} Electronic (top) and phonon (bottom) band structures of Ag.

\medskip
\noindent\textbf{Figure~\ref{fig:bs_al}.} Electronic (top) and phonon (bottom) band structures of Al.

\medskip
\noindent\textbf{Figure~\ref{fig:bs_libept2}.} Electronic (top) and phonon (bottom) band structures of LiBePt$_2$.

\medskip
\noindent\textbf{Figure~\ref{fig:bs_nbw2}.} Electronic (top) and phonon (bottom) band structures of NbW$_2$.

\end{document}